\documentclass[final,3p,times]{elsarticle}
\usepackage{amssymb}
\usepackage{amsmath}
\journal{Physica A}

\def\Eq#1{Eq.~(\ref{#1})}
\def\Eqs#1{Eqs.~(\ref{#1})}

\def\no{\nonumber\\}
\def\r{\rangle\!\rangle}

\def\>{\rangle}
\def\<{\langle}

\def\adg{a^\dagger}
\def\tr{\text{tr}}
\def\dg{\dagger}
\def\Kt{\widetilde{K}}
\def\Adg{A^\dg}
\def\At{\widetilde{A}}
\def\Atdg{{\widetilde{A}}^\dg}

\def\J{\mathbb{J}}

\def\kap{\kappa}
\def\w{\omega}
\def\del{\delta}
\def\gam{\gamma}

\def\al{\alpha}
\def\bt{\beta}
\def\th{\theta}
\def\g{\text{G}}

\def\xh{\hat{x}}
\def\ph{\hat{p}}

\def\xt{\widetilde{x}}

\def\d{\partial}
\def\T{\text{T}}
\def\h{h}
\def\u#1{\mathbf{#1\!}\,	}
\def\KL{\text{KL}}
\def\CL{\text{CL}}
\def\HPZ{\text{HPZ}}

\def\s{\mbox{\boldmath$\sigma$}}
\def\bbt{\mbox{\boldmath$\beta$}}

\begin{document}

\begin{frontmatter}

\title{Symmetry of bilinear master equations for a quantum oscillator}

\author[1]{B. A. Tay}
\ead{BuangAnn.Tay@nottingham.edu.my}
\address[1]{Foundation Studies, Faculty of Engineering, The University of Nottingham Malaysia Campus, Jalan Broga, 43500 Semenyih, Selangor, Malaysia}

\date{\today}

\begin{abstract}
We study the most general continuous transformation on the generators of bilinear master equations of a quantum oscillator. We find that transformation operators that preserve the hermiticity of density operators and conserve the probability of reduced dynamics should be adjoint-symmetric, and they are not limited to the pure product of unitary operators in the bra and ket space but could be a mixture of them. We need to include the more general transformation operators to explore the full symmetry of generic reduced dynamics. We discuss how the operators are related to those considered in previous works, and illustrate how they leave the reduced dynamics form invariant, or map one into the other. The positive semidefinite requirement on the density operator can be imposed to give a valid range of transformation parameters.
\end{abstract}

\begin{keyword}
Symmetry \sep Liouville space \sep Reduced dynamics \sep Open quantum systems
\PACS 05.70.Ln
\end{keyword}

\end{frontmatter}

\section{Introduction}

The time evolution of isolated quantum systems is initiated by unitary transformation \cite{Dirac} $\rho_\text{tot}\rightarrow u_\text{tot}\rho_\text{tot}u_\text{tot}^\dg$. When we are interested in the behaviour of a subsystem within the isolated system, we usually treat the rest of the degrees of freedom as the environment and take an average over them \cite{Sudarshan61,Gardiner,Breuer}. If the system belongs to the class of large Poincar\'e systems that is non-integrable \cite{Petrosky91,Petrosky97}, i.e., it possesses unstable states or resonances, the subsystem will exhibit an irreversible evolution endowed with a semigroup property, which eventually evolves into a stationary state. The evolution of the subsystem is described by a non-unitary transformation $\rho=\tr_\text{env} \rho_\text{tot} \rightarrow \sum_i u_i \rho u_i^\dg$ \cite{Sudarshan61,Kraus83} that is not factorizable \cite{Prigogine73,Gonzalo01,Sungyun03}.

In analogy to the case of time evolution, we can consider a symmetry $s_\text{tot}$ possessed by the  original isolated system as a whole $\rho_\text{tot}\rightarrow s_\text{tot} \rho_\text{tot} s^\dg_\text{tot}$. After the environment degrees of freedom are averaged over, we are left with a reduced or partial symmetry on the subsystem. Similar to the evolution of the subsystem, the left over symmetry on the subsystem in general need not be unitary or factorizable, though it must be consistent with the reduced dynamics and its semigroup property.

In this work, we investigate the most general continuous symmetry on the subsystem of an oscillator with bilinear generators. Previous studies on the symmetry of the reduced dynamics mainly focused on transformation that is unitary and factorizable \cite{Barnett85,Ekert90,Ban92,Ban93} using the symplectic group in two dimensions or the SU(1,1) group, whereas the possibility of a more general symmetry was not considered. Even though the general form of the reduced dynamics for an oscillator that satisfies quantum Gauss Markov process had been derived \cite{Talkner81}, and the solutions of bilinear master equations of an oscillator under different conditions and models of noise \cite{Weiss,Hu11,Grabert97} were obtained, but the symmetry aspects of the reduced dynamics and the solutions were not investigated.

To be consistent with the reduced dynamics, we require the transformation to satisfy a few properties. (1) They preserve the commutation relations of the creation and annihilation operators of the oscillator. Hence, they must belong to the symplectic group in four dimensions with ten generators \cite{Gilmore,Simon88}. (2) They preserve the hermiticity of the reduced density operators. Consequently, they must be adjoint-symmetric \cite{Prigogine73}, and this fact rules out some of the unitary transformations that are not factorizable. (3) They conserve the probability of the reduced dynamics. Therefore, we require their generators to have zero trace with the reduced density operator. This requirement limits the valid number of the generators of transformation to seven. (4) The positive semidefiniteness \cite{Kossa76,Lindblad76,Pechukas94,Shaji05} requirement is imposed as an extra condition that further restricts the parameter spaces of the transformation. (5) As a result, the transformed dynamics satisfies the semigroup property. We then show how the transformation leaves a few typical reduced dynamics \cite{Kossa76,Lindblad76,CL83,HPZ92} form invariant, and how it maps them into one another.

\section{Requirements on the transformation}
\label{SecReqTransf}

Before we construct the transformation operators, we first clarify conditions that must be fulfilled by the transformation so that it is consistent with the properties of the reduced dynamics stated in the Introduction \cite{Sudarshan61}. We consider a time-independent transformation operator $S$ on the density operator
\begin{align}   \label{rhoprime}
    \rho'(t)=S\rho(t)
\end{align}
that has an inverse. The hermiticity condition is usually stated as $S\rho^\dg=(S\rho)^\dg$. It can also be written as $(S\rho)^\dg=\rho^\dg S^\dg\equiv\widetilde{S}\rho^\dg$ by means of the association operation \cite{Prigogine73} denoted by $\sim$, discussed in \ref{AppOpSop}. This operation is equivalent to the tilde conjugation introduced in thermofield dynamics \cite{Umezawa75,Umezawa75b}. As a result, $S$ should be adjoint-symmetric,
\begin{align} \label{Ut=U}
    \widetilde{S}&=S\,.
\end{align}	
Likewise, the generator of the reduced dynamics $K$ in the equation of motion $\d\rho/\d t=-K(t)\rho$ is adjoint-symmetric as well
\begin{align}   \label{KtildeK}
    \Kt(t)=K(t)\,,
\end{align}
and so does the time evolution operator, see the proof in \ref{AppTimeEvo}.

As an observable, $K(t)$ then transforms under a similarity transformation
\begin{align} \label{Kprime}
    K'(t)\equiv SK(t)S^{-1}\,.
\end{align}
If $S$ and $K(t)$ are both adjoint-symmetric, it follows that the transformed generator is adjoint-symmetric too,
\begin{align}   \label{Kt'=K'}
    \Kt'(t)&=\big(SK(t)S^{-1}\big)^{\widetilde{\,\,}}=\widetilde{S}\Kt(t)\widetilde{S}^{-1}=K'(t)\,,
\end{align}
where we use the second property of the association operation in \Eq{tilde}. For time-independent transformation, $\rho'$ obeys a similarly related equation of motion $\d\rho'/\d t=-K'(t)\rho'$.\footnote{If we were to consider time-dependent transformation, additional terms would have to be added to the transformed equation of motion to take into account the time-dependence of the parameters of transformation. We will not consider time-dependent transformation in this work.}

In order to preserve the normalization condition $\tr\rho=1$ as in $\d(\tr\rho)/\d t=0$, the generator must have zero trace with the reduced density operator
\begin{align}   \label{trKrho0}
    \tr(K(t)\rho)=0\,.
\end{align}
Similarly, the transformed generator must also satisfy
\begin{align} \label{trK'=0}
    \tr\big(K'(t)\rho'\big)=0\,,
\end{align}
so that $\d(\tr\rho')/\d t=0$.

On the other hand, implementing the positive semidefiniteness condition on $\rho'$ is not as straight forward as the other requirements. We know that generators that are completely positive \cite{Kossa76,Lindblad76} preserve the positive semidefiniteness of $\rho$ throughout the course of their evolutions. However, this is not the case with other forms of generators \cite{Pechukas94,Shaji05}. The violations of positive time evolution can be avoided through a slippage of initial conditions \cite{Suarez92,Gaspard99}. On the other hand, if we know the explicit solutions of the reduced dynamics, we can impose the positive semidefiniteness requirement on the reduced density operators directly \cite{Shaji05}. This leads to the positive domain \cite{Shaji05} of the reduced dynamics that ensures the positive time evolution of the subsystem.

Since the transformation is invertible, the evolution operator of the transformed dynamics inherits the semigroup property from the original dynamics,
\begin{align} \label{V't}
    V'(t_1+t_2)\equiv SV(t_1+t_2)S^{-1}=SV(t_1)S^{-1}SV(t_2)S^{-1}=V'(t_1)V'(t_2)\,, \qquad t_1, t_2\geq 0\,.
\end{align}

\section{General transformation operators for an oscillator}
\label{SecGenFormRedDyn}

In this section, we construct transformation operators for an oscillator that satisfy the requirements stated in Section \ref{SecReqTransf}. The four basic operators of an oscillator in the reduced space are
\begin{align} \label{AA}
    A\equiv a\times 1 \,, \qquad \Adg\equiv \adg\times 1\,,\qquad
    \At\equiv 1\times \adg\,, \qquad \At^\dg\equiv 1\times a\,,
\end{align}
where $\adg$ and $a$ are the creation and annihilation operators of the oscillator, respectively. From the commutation relation $[a, \adg]=1$, it follows from the multiplication rule of operators (see \ref{AppOpSop}) that
\begin{align}   \label{commAA}
    [A,\Adg]=1\,, \qquad [\At, \Atdg]=1\,,
\end{align}
while other commutators vanish.

There are ten bilinear operators that can be constructed out of the four basic operators. They generate the symplectic group in four dimensions \cite{Gilmore08}. By taking specific complex combinations of the bilinear operators, we obtain the following ten generators,
\begin{subequations}
\begin{align} \label{iL0A}
    iL_0&\equiv\frac{i}{2}(\Adg A-\Atdg\At)\,, \qquad\qquad\qquad\qquad\qquad\qquad\qquad O_0\equiv\frac{1}{2}(\Adg\Atdg-A\At)\,,\\
    iM_1&\equiv\frac{i}{4}(\Adg\Adg+AA-\Atdg\Atdg-\At\At)\,,\qquad\qquad\qquad\qquad
    iM_2\equiv\frac{1}{4}(\Adg\Adg-AA+\Atdg\Atdg-\At\At)\,,\\
    O_+&\equiv\frac{1}{2}(\Adg\Atdg+A\At-\Adg A-\Atdg\At-1)\,,\qquad\qquad\qquad O_-\equiv\frac{1}{2}(\Adg\Atdg+A\At+\Adg A+\Atdg\At+1)\,,\\
    L_{1+}&\equiv\frac{1}{4}(2\Adg\At+2A\Atdg-\Adg\Adg-AA-\Atdg\Atdg-\At\At)\,,\qquad
    L_{1-}\equiv\frac{1}{4}(2\Adg\At+2A\Atdg+\Adg\Adg+AA+\Atdg\Atdg+\At\At)\,,\\
    L_{2+}&\equiv-\frac{i}{4}(2\Adg\At-2A\Atdg-\Adg\Adg+AA+\Atdg\Atdg-\At\At)\,,\qquad
    L_{2-}\equiv-\frac{i}{4}(2\Adg\At-2A\Atdg+\Adg\Adg-AA-\Atdg\Atdg+\At\At)\,.\label{L2-A}
\end{align}
\end{subequations}
They have the desired properties discussed in the next section.

\subsection{Properties of generators}
\label{SecTraceBasisProp}

The generators can be divided into three subsets,
\begin{align}   \label{V0}
    \J_0=\{iL_0, iM_1, iM_2, O_0\}\,,\qquad
    \J_+=\{O_+,L_{1+},L_{	2+}\}\,,\qquad
    \J_-=\{O_-,L_{1-},L_{2-}\}\,.   
\end{align}

\begin{enumerate}
\item They are adjoint-symmetric, $\widetilde{J}=J \in \J_0, \J_\pm$.

\item They form a complete and orthogonal set of basis under the Hilbert-Schmidt norm \cite{Nielsen} in the four dimensional matrix representation of the generators obtained in \ref{AppMrepres}.

\item Generators from $\J_0$ and $\J_+$ have zero traces with arbitrary density operators,
\begin{align}   \label{J0+rho}
    \tr (iL_0\rho)&=0=\tr (iM_1\rho)=\tr (iM_2\rho)=\tr \big[(O_0-I/2)\rho\big]=\tr(O_+\rho)=\tr(L_{1+}\rho)=\tr(L_{2+}\rho) \,,
\end{align}
where $I\equiv 1\times 1$ is the identity superoperator and we have used the cyclic invariance of trace to arrive at \Eq{J0+rho}. In contrast, the generators from $\J_-$ give
\begin{align}   \label{J0-rho}
    \tr (L_{1-}\rho)&=\tr\big[(\adg\adg+aa)\rho\big]\,, \qquad \tr(L_{2-}\rho)=-i\,\tr\big[(\adg\adg-aa)\rho\big]\,, \qquad  \tr(O_-\rho)=\tr\big[(\adg a+a\adg)\rho\big]\,.
\end{align}
No linear combination of these operators with others can have zero trace with arbitrary density operators.

\item Generators from the $\J_0$ and $\J_+$ spaces are closed under the commutator brackets. The commutation relations of the generators are summarized in Table \ref{commV} using the notation
\begin{align}
    \begin{array}{c||c}
            & J'  \\
           \hline\hline
           J & J''
          \end{array}
    \qquad \leftrightarrow \qquad [J,J']=J'' \,.\nonumber
\end{align}

\begin{table}[t]
\center
\begin{tabular}{c||cccc|ccc|ccc}
                 & $iL_0$ & $iM_1$ & $iM_2$ & $O_0$ & $O_+$ & $L_{1+}$ &
                  $L_{2+}$ & $O_-$ & $L_{1-}$& $L_{2-}$ \\
                  \hline\hline
                 $iL_0$ & 0 & $-iM_2$ & $iM_1$ & 0 & 0 & $-L_{2+}$ & $L_{1+}$ & 0 & $-L_{2-}$ & $L_{1-}$ \\
                 $iM_1$ & $iM_2$ & 0 & $iL_0$ & 0 &  $L_{2+}$ & 0 & $O_+$ &  $L_{2-}$ & 0 & $O_-$\\
                 $iM_2$ &  $-iM_1$ &  $-iL_0$ & 0 & 0 &  $-L_{1+}$ &  $-O_+$ & 0 &  $-L_{1-}$ &  $-O_-$ & 0 \\
                 $O_0$ & 0 & 0 & 0 & 0 & $O_+$ &  $L_{1+}$ &  $L_{2+}$ &  $-O_-$ &  $-L_{1-}$ & $-L_{2-}$ \\
                 \hline
                 $O_+$ & 0 &  $-L_{2+}$ &  $L_{1+}$ &  $-O_+$ & 0 & 0 & 0 &  $-2O_0$ &  $-2iM_2$ &  $2iM_1$ \\
                 $L_{1+}$ &  $L_{2+}$ &  0 & $O_+$ & $-L_{1+}$ & 0 & 0 & 0 & $2iM_2$ & $2O_0$ & $2iL_0$ \\
                 $L_{2+}$ & $-L_{1+}$ & $-O_+$ & 0 & $-L_{2+}$ & 0 & 0 & 0 & $-2iM_1$ & $-2iL_0$ & $2O_0$ \\
                 \hline
                 $O_-$ & 0 & $-L_{2-}$ & $L_{1-}$ & $O_-$ & $2O_0$ & $-2iM_2$ & $2iM_1$ & 0& 0 & 0 \\
                 $L_{1-}$ & $L_{2-}$ & 0 & $O_-$ & $L_{1-}$ & $2iM_2$ & $-2O_0$ & $2iL_0$ & 0 & 0 & 0 \\
                 $L_{2-}$ & $-L_{1-}$ &$-O_-$ & 0 & $L_{2-}$ & $-2iM_1	$ & $-2iL_0$ & $-2O_0$ & 0 & 0 & 0 \\
\end{tabular}
\caption{Commutation relations of the generators.}
\label{commV}
\end{table}
They have the structure
\begin{align}   \label{Jcomm}
    [\J_0,\J_0]\subseteq \J_0\,,   \qquad   [\J_\pm,\J_\pm]=0\,,\qquad %
    [\J_0, \J_\pm]\subseteq \J_\pm\,,  \qquad   [\J_+, \J_-]\subseteq \J_0\,.
\end{align}
In particular, $[O_0, \J_0]=0$, and $[O_0, \J_\pm]=\pm \J_\pm$.

\item Generators in $\J_0$ are anti-hermitian and anti-symmetric,
\begin{align}   \label{antihermi}
    J^\dg=-J\,, \qquad J^\T=-J\,,\qquad J\in \J_0\,.
\end{align}
Generators in $\J_\pm$ are hermitian and symmetric,
\begin{align}   \label{hermi}
    J^\dg=J\,, \qquad J^\T=J\,,\qquad J\in \J_\pm\,.
\end{align}
\end{enumerate}

\subsection{Adjoint-symmetric transformation operators}

Based on the trace property 3 of the generators, and in view of the requirement in \Eq{trKrho0}, generators from the $\J_0$ and $\J_+$ spaces are consistent with the probability conserving nature of the reduced dynamics. Consequently, the generic form of the generators of the reduced dynamics for an oscillator is
\begin{align}   \label{K}
    K
    =\h_0iL_0+\h_1iM_1+\h_2iM_2+g_0(O_0-I/2)+g_+O_++g_1L_{1+}+g_2L_{2+}\,,
\end{align}
where $\h_j$ and $g_j$ are real coefficients to ensure the adjoint-symmetry of $K$. We have suppressed the time-dependence on the coefficients to simplify the expression. The $iL_0$ component generates the free evolution of the oscillator, whereas $iM_1$ renormalizes the frequency of the oscillator, see the discussion in Section \ref{SecSymRD}. They generate the unitary evolution of the reduced dynamics. We note that an equivalent form of reduced dynamics for quantum Gauss Markov process was derived in Ref.~\cite{Talkner81}.

We form the transformation operators by exponentiating the generators,
\begin{align} \label{UV}
    S(\al)\equiv\exp\left( \al J\right)\,,
\end{align}
where $\al$ is real to ensure the adjoint-symmetry of $S$. The inverse operator is
\begin{align} \label{invS}
    S^{-1}(\al)=S(-\al)\,.
\end{align}

To ensure that the transformed reduced dynamics \eqref{Kprime} again belongs to the $\J_0$ and $\J_+$ spaces, we restrict the generators of $S$ to the $\J_0$ and $\J_+$ spaces only, since they are closed under the commutator brackets by property 4 in Section \ref{SecTraceBasisProp}. As a result, conditions \eqref{Kt'=K'} and \eqref{trK'=0} are satisfied. Transformation operators with generators from the $\J_-$ space are not allowed because they bring $K'$ out of the $\J_0$ and $\J_+$ spaces, thereby violating the probability conservation requirement of the reduced dynamics. Once we impose the positive semidefiniteness condition on the reduced density operator, there will be further restrictions on the range of the transformation parameters. The semigroup property of the reduced dynamics then follows as discussed at the end of Section \ref{SecReqTransf}.

The generators of $S$ consist of a linear combination of the bilinear generators over complex numbers \eqref{iL0A}-\eqref{L2-A}. $S$ leaves the commutation relations \eqref{commAA} invariant because they are elements of the symplectic group \cite{Gilmore}. Consequently, we find that transformation operators that are consistent with the reduced dynamics form a subgroup of the complex symplectic group in four dimensions. To our knowledge, the basis \Eqs{iL0A}-\eqref{L2-A} has not been introduced in the literature, and the subgroup generated by $\J_0$ and $\J_+$ has not been studied before.

We note that we have not incorporated the operator $-I/2$ into the definition of $O_0$, because $I$ is not a generator of the symplectic group. Incorporating it into $O_0$ will produce an operator $\exp[\al(O_0-I/2)]$ that gives rise to an overall factor $\exp(-\al/2)$ on the left hand side of the quadratic condition satisfied by symplectic group \eqref{SSTbtSS}, which is specialized to the four dimensional matrix representation of the generators. Hence, it violates the quadratic condition and is not incorporated into the definition of $O_0$.

\subsection{Relation to other transformations}
\label{SecRelOther}

The anti-hermiticity of the generators in $\J_0$ \eqref{antihermi} implies that $\exp(\th iL_0)$, $\exp(\phi iM_1)$ and $\exp(\phi iM_2)$ are unitary, $S^\dg=S^{-1}$. Moreover, they are factorizable, i.e., they can be written in the form $u_j\times u_j^\dg$, $j=0, 1,2 $, respectively, where $u_0=\exp(i\th \adg a/2)$, $u_1=\exp[i\phi (\adg\adg+ aa)/4]$, $u_2=\exp [\phi (\adg\adg- aa)/4]$, and $u_j^\dg=u_j^{-1}$. These factorizable operators maintain the purity of pure states. The generators are isomorphic to those of the canonical transformation discussed in Ref.~\cite{Ekert90}. In fact, $u_1$ and $u_2$ are the squeezing operators \cite{Gardiner}.

On the other hand, $\exp(\al O_0)$ is an example of a unitary but not factorizable operator. It brings the vacuum state (pure state), $|0;0\r=|0\>\<0|$, into the Gibbs states (mixed states) \cite{Tay11},
\begin{align}   \label{GOG}
    \rho_\g&\equiv e^{\al O_0}\vert0;0\r=\sqrt{1-e^{-2\w_0/kT}}\sum_{n=0}^\infty e^{-n\w_0/kT}\vert n;n\r\,,
\end{align}
where $\w_0$ is the natural frequency of the oscillator, and the transformation parameter $\al$ is related to the temperature of the environment $T$ by $\exp\al=\coth(\w_0/2kT)$. The $\exp(\al O_0)$ is the generalized Bogoliubov transformation operator used in the studies of thermofield dynamics \cite{Umezawa75,Umezawa75b,Barnett85,Santana06a}. It also generates the thermal symmetry \cite{Tay07} to be discussed in Section \ref{SecFormInv}.

The hermiticity of the generators in $\J_+$ \eqref{hermi} implies that $\exp(\al O_+)$, $\exp(\al L_{1+})$, and $\exp(\al L_{2+})$ are not unitary. Furthermore, they cannot be factorized into the form $u\times v$. Though we can form operators that are unitary out of them, for instance, $\exp(i\al O_+), \exp(i\al L_{1+})$ and $\exp(i\al L_{2+})$, but these operators are not adjoint-symmetric and hence cannot be valid transformation operators on the reduced dynamics, see also the comments in \ref{AppRelU}. The set of generators of the SU(1,1) group used in Ref.~\cite{Ban92} are isomorphic to $O_0, O_+$ and $O_-$. The trace preserving requirement is not manifest using this set of generators. Later in Section \ref{SecFormInv} and \ref{SecBetwRedDyn}, we will show that $\exp(\al O_+)$, $\exp(\al L_{1+})$, and $\exp(\al L_{2+})$ are needed to construct the symmetry of more general reduced dynamics than those considered in Refs.~\cite{Ekert90,Ban92}, for example.

A generic generator of the reduced dynamics that annihilates the vacuum state has the form
\begin{align} \label{KV}
    K_V\equiv h_0 iL_0+g_0(O_0-I/2-O_+)
    +h_1(iM_1-L_{2+})+h_2(iM_2+L_{1+})\,,
\end{align}
because $iL_0$ and the combinations of generators in the round brackets separately annihilate the vacuum state. Since the Gibbs states are generated from the vacuum state by $\exp(\al O_0)$ \eqref{GOG}, reduced dynamics with the Gibbs states as their stationary states have the generic form
\begin{align}   \label{KbGibbs}
    K'_V&\equiv e^{\al O_0}K_V e^{-\al O_0}
    =h_0iL_0+ h_1iM_1+ h_2iM_2+ g_0(O_0-I/2)
    -2bg_0O_+ + 2bh_2L_{1+} -2bh_1L_{2+}\,,
\end{align}
where $2b=\exp\al$.

The displacement operators that generate the coherent states \cite{Klauder68,Gardiner,Ban92} and the coherent thermofield \cite{Barnett85} do not appear in the above analysis because they are constructed by exponentiating operators linear in the four basic operators \eqref{AA}. Since they are not the main objects of our investigation, we discuss them in \ref{AppDop} for completeness. Reduced dynamics with coherent states as its stationary states is obtained using the displacement operators in \ref{AppDop}.

\section{Symmetry of reduced dynamics}
\label{SecSymRD}

In this section, we consider transformation that leaves the reduced dynamics form invariant, or maps it into the other reduced dynamics. The effects of the transformation operators are best illustrated with some concrete examples. Three often used reduced dynamics in the studies of quantum Brownian motion \cite{CL83,HPZ92}, quantum decoherence \cite{Giulini}, quantum information \cite{Nielsen}, quantum optics \cite{Gardiner}, and etc, are the Kossakowski-Lindblad (KL) equation \cite{Lindblad76,Kossa76}, the Caldeira-Leggett (CL) equation \cite{CL83}, and the Hu-Paz-Zhang (HPZ) equation \cite{HPZ92} where we consider its markovian regime \cite{Tay06}. Their generators are
\begin{align}   \label{KKL}
    K_\KL(b)&\equiv 2\w_0 iL_0+\gam (O_0-I/2)-2\gam b O_+\,,\\
    K_\CL(b)&\equiv 2\w_0 iL_0+\gam(O_0-I/2-iM_2)-2\gam b(O_++L_{1+})\,,\label{KCL}\\
    K_\HPZ(b,d)&\equiv 2\w_0 iL_0+\gam(O_0-I/2-iM_2)-2\gam b(O_++L_{1+})-dL_{2+}\,,\label{KHPZ}
\end{align}
respectively. In the equations, we use the units $\hbar=1$, $\w_0$ is the natural frequency of the oscillator, $\gam$ is the relaxation or damping rate, and $b$ is the thermal parameter related to the temperature of the reservoir $T$ by
\begin{align}   \label{b}
    b&=\frac{1}{2} \coth \frac{\w_0}{2kT}\,.
\end{align}
All the coefficients are real and $\w_0, \gam, b$ are positive. The CL equation was obtained in the high temperature limit \cite{CL83} where $b$ goes into $k T/\w_0$. Here we treat $d$ in the HPZ equation as an independent coefficient.

We have listed the effects of the transformation operators on the generic generator of reduced dynamics in \ref{AppTable}. To study the symmetry of more general reduced dynamics, such as $K_\CL$ and $K_\HPZ$ that contain the $L_{1+}$ and $L_{2+}$ components, we have to include transformation that is not unitary and not factorizable. An important observation we make from the list is that the relaxation rate cannot be altered by any of the operators.

In the generators of the reduced dynamics, there is no loss of generality that we have omitted the $iM_1$ component. This is because $2\w_0 iL_0+h_1 iM_1$ that generates the unitary evolution of the oscillator can always be diagonalized by $\exp(i\phi M_2)$ into $2\w'_0 iL_0$, cf. \Eq{U2K}, where
\begin{align}   \label{w0renorm}
    \w'_0= \sqrt{\w_0^2-h_1^2/4}\,, \qquad
    \tanh\phi=\frac{h_1}{2\w_0}\,.
\end{align}
This also shows that the $iM_1$ component renormalizes the frequency of the oscillator.

\subsection{Form invariant transformation}
\label{SecFormInv}

Now we consider symmetry that leaves the reduced dynamics form invariant, while keeping the frequency of the oscillator unchanged.

\begin{enumerate}
\item The operator $\exp(\al O_0)$ keeps the form of $K_\KL$ and $K_\CL$ invariant,
\begin{align}   \label{KLinv}
    e^{\al O_0}K_{\KL\,\text{or}\, \CL}(b) e^{-\al O_0} =K_{\KL\,\text{or}\, \CL}(b')\,.
\end{align}
It introduces a dilation factor $\exp\al$ to the coefficients of the $O_+, L_{1+}, L_{2+}$ components, cf. \Eq{U0K}. This generates the so-called thermal symmetry in the reduced dynamics \cite{Tay07}, where the system is brought to a different temperature determined by
\begin{align}   \label{b'O0}
    b'= be^\al\,.
\end{align}
In $K_\HPZ$, $d$ is mapped to $d'=d\exp\al$. Since $K_\HPZ$ has two coefficients, $\exp(\al O_0)$ does not exhaust the full symmetry of $K_\HPZ$, which is considered in case 3 below.

\item A translated form of thermal symmetry is initiated by $\exp(\bt O_+)$ on $K_\CL$, cf. \Eq{U1+K},
\begin{align}   \label{CLinv}
    &e^{\bt O_+}K_\CL(b) e^{-\bt O_+}=K_\CL(b')\,,
\end{align}
where the thermal parameter is translated into
\begin{align}   \label{btranslate}
    b'=b+\bt/2\,.
\end{align}

\item Since $K_\HPZ$ has two coefficients, the general symmetry can be described by at least two parameters. We find that it is kept form invariant through
\begin{align}   \label{KHPZinv}
    e^{\phi i M_2}e^{\xi (O_++L_{1+})}e^{\phi (O_0-iM_2)}K_\HPZ(b,d) e^{-\phi (O_0-iM_2)}e^{-\xi (O_++L_{1+})} e^{-\phi i M_2}
    =K_\HPZ(b',d')\,,
\end{align}
where
\begin{align}   \label{b'HPZ}
    b'=be^\phi+\xi e^{-\phi}\,, \qquad \frac{d'}{2\w_0}=\frac{d}{2\w_0}e^\phi-\xi e^{-\phi}\,.
\end{align}
The rotation induced by $\exp(\phi iM_2)$ is required to keep the natural frequency $\w_0$ invariant.

\end{enumerate}

\subsection{Positive domain of parameters}
\label{SecPosDomain}

We can determine the valid range of the transformation parameters by investigating the domain of the parameters that maintains the positive semidefiniteness of the density operator. As a specific example, we consider the stationary states of the reduced dynamics. For this purpose, it is more convenient to work in the position representation described in \ref{AppQr}.

The stationary state of $K_\KL, K_\CL$ and $K_\HPZ$ in the position representation is
\begin{align}   \label{rhoequi}
\rho(b,d)=N\exp\left(-\frac{Q^2}{2b+d/\w_0}-b\frac{r^2}{2}\right)\,,
\end{align}
where $d=0$ for $K_\KL, K_\CL$, and $N$ is an unimportant normalization constant.

As explained in Ref.~\cite{Simon88}, the necessary and sufficient conditions for the positive semidefiniteness of single mode density operator with the Gaussian profile $\exp[-2\mu Q^2-i\kap Qr-(\mu+\nu)r^2/2]$ is $\mu>0$ and $\nu\geq 0$. When we apply these conditions to $\rho(b,d)$ \eqref{rhoequi}, we obtain
\begin{align} \label{+ineq}
    2b+\frac{d}{\w_0}>0\,, \qquad
    2b\left(2b+\frac{d}{\w_0}\right)\geq 1\,,
\end{align}

When $d=0$, as in $K_\KL$ and $K_\CL$, \Eq{+ineq} reduces to a single inequality $b\geq 1/2$, which is equivalent to $T\geq 0$. When the density operator is mapped into $\rho'(b',d')=S\rho(b,d)$, $b', d'$ have to satisfy a set of inequalities similar to \Eq{+ineq} to ensure the positive semidefiniteness of $\rho'$.

We will now consider the positive domain of each of the three transformation operators considered in Section \ref{SecFormInv}.

\begin{enumerate}
\item   The positive semidefiniteness condition on the transformed density operator $\rho'_{\KL\,\text{or}\, \CL}=e^{\al O_0}\rho_{\KL\,\text{or}\, \CL}$ gives $b'=b\exp\al\geq 1/2$. This restricts the parameter of transformation to $\al \geq \ln 2b$.

\item Imposing the positive semidefiniteness condition on $\rho'_\CL=e^{\bt O_+}\rho_\CL$ gives $b'=b+\bt/2\geq 1/2$, which yields the restriction $\bt \geq -(2b-1)$.

\item   For $\rho'_\HPZ=e^{\phi i M_2}e^{\xi (O_++L_{1+})}e^{\phi (O_0-iM_2)}\rho_\HPZ$, substituting \Eq{b'HPZ} into \Eq{+ineq} gives
\begin{align}
    e^\phi>0\,, \qquad \xi\geq \frac{2}{2b+d/\w_0}-2be^{2\phi}\,.
\end{align}
It follows that all real $\phi$ are permitted, and the range of $\xi$ is restricted by the second inequality.

\end{enumerate}
In this way, we learn how the positive semidefinite requirement on the density operators further constraints the valid range of the transformation operators, on top of the restrictions imposed by the hermiticity and probability conservation requirements.

\subsection{Transformation between reduced dynamics}
\label{SecBetwRedDyn}

The non-factorizable operators $\exp(\al L_{1+}), \exp(\al L_{2+})$ must be used to initiate the transformation between $K_\KL$, $K_\CL$ and $K_\HPZ$. For example, we find that $K_\KL$ is mapped into $K_\CL$ through
\begin{align}
K_\CL(\w_0',b')=e^{\th iM_1}e^{\eta L_{2+}}K_\KL(\w_0,b)e^{-\eta L_{2+}}e^{-\th iM_1}\,,
\end{align}
provided that the transformation parameters take the values $\eta=-2b\tanh\th$ and $\sinh\th=-\gam/(2\w_0)$. From the observation made at the end of \ref{AppTable}, the natural frequency of the oscillator cannot be kept constant in this case. The coefficients are related by $\w'_0=\w_0 \cosh\th$, and $b'=b/\cosh\th$. If we apply the positive semidefiniteness condition on the stationary state, we find that $\th$ and $\eta$ must satisfy the constraints $\cosh \th\leq 2b$ and $\eta \geq \gam/(2\w_0)$, respectively.

$K_\CL$ can be mapped into $K_\HPZ$ through a single transformation
\begin{align}   \label{CLHPZ}
    K_\HPZ( b', d')
    &= e^{\zeta L_{1+}}K_\CL( b) e^{-\zeta L_{1+}}\,,
\end{align}
where $b'=b+\zeta/2$, and $d'=-2\w_0\zeta$. The positive semidefiniteness condition on the transformed stationary state constrains $\zeta$ to the range $-\sqrt{4b^2-1}\leq\zeta\leq \sqrt{4b^2-1}$.

Since the inverse of the transformed operators \eqref{invS} exists, we have effectively established the maps between $K_\KL, K_\CL$, and $K_\HPZ$.

If two reduced dynamics are physically equivalent, the expectation values of observables before and after the transformation should be the same. In \ref{AppExpValue}, we show that the expectation values of observables remain invariant if the transformation is unitary, regardless of whether it is factorizable or not. Consequently, $K_\KL, K_\CL$ and $K_\HPZ$ are not physically equivalent, i.e., the mappings between them necessarily involve the non-unitary operators $\exp(\al L_{1+})$ and $\exp(\al L_{2+})$, because $K_\CL$ and $K_\HPZ$ contain the $L_{1+}$ and $L_{2+}$ components. For the form invariant transformation discussed in Section \ref{SecFormInv}. Though $\exp(\al O_0)$ is unitary, but because of its non-factorizable nature, it maps the subsystem to a different temperature. As a result, although the original and the transformed subsystems have the same dynamical behaviour, but they are not physically the same. We emphasize that the mappings we discuss cannot be achieved by the unitary and factorizable operators $\exp(\theta iL_0), \exp(\phi iM_1)$ and $\exp(\phi iM_2)$ alone.

\section{Conclusion}
\label{SecConclusion}

We find that the unitary transformation that is factorizable in reversible systems does not exhaust the full symmetry of the generator of reduced dynamics with irreversible time evolution. Transformation operators that are consistent with the reduced dynamics should be adjoint-symmetric. They not only include the usual unitary operators that are factorizable, but extend beyond them. The extended operators mix operators in the bra and ket space, or the Liouville space. Consequently, their effects cannot be reproduced by ordinary operators in the Hilbert space. We can also identify extended unitary operators that are inconsistent with the reduced dynamics if they are not adjoint-symmetric.

In this work, we have only considered kinematical aspects of the symmetry for the subsystem of a quantum oscillator, such as the transformation on the generators of its reduced dynamics and their stationary states. To consider dynamical aspects of the symmetry that involve the time evolution of the subsystem, explicit solutions to the equation of motion are required. For this purpose, the set of adjoint-symmetric generators introduced here enables us to analyze the time evolution operator systematically. As already discussed in Section \ref{SecReqTransf}, the generators of the Caldeira-Leggett and the Hu-Paz-Zhang equation do not guarantee a positive time evolution of the reduced dynamics. Therefore, the positive conditions obtained for the stationary states in Section \ref{SecPosDomain} are not sufficient to ensure the positive time evolution of the subsystem. Sufficient condition can be studied once the solutions to the equation of motion are known. It is also interesting to investigate whether the formulation can be implemented in finite-level systems. We will discuss the other aspects of the symmetry in a future work.

\hfill

\noindent \textbf{Acknowledgments}

\hfill

We thank the late Professor Syed Twareque Ali and Professor Hishamuddin Zainuddin for interesting discussions during the Expository Quantum Lecture Series 8 (EQuaLS8) in 2016. We thank Professor Kazunari Hashimoto and Professor Chikako Uchiyama for their hospitality, Professor Gonzalo Ordonez and Professor Sungyun Kim for interesting discussions during the International Symposium on Foundation of Quantum Transport in Nano Science in 2015. We also thank Professor Hisao Hayakawa and Professor Tomio Petrosky for their suggestions to generalize the previous work in Ref.~\cite{Tay07} during the Yukawa International Program for Quark-Hadron Sciences (YIPQS) in 2008 that resulted in this work. This work is supported by the Ministry of Higher Education Malaysia (MOHE) under the Fundamental Research Grant Scheme (FRGS), Project No.~FRGS/2/2014/ST02/UNIM/02/1.

\appendix

\section{Operations on superoperators}
\label{AppOpSop}

Before we introduce the notion of adjoint-symmetry, we need to first introduce a few operations on superoperators \cite{Prigogine73}. We use capital letters to denote superoperators. We define two superoperators $X\equiv x_1\times x_2$ and $Y\equiv y_1 \times y_2$, where $x_1, y_1$ and $x_2, y_2$ are operators that act on the ket- and bra-space, respectively. Superoperators act on an operator $\rho$ from the left as $X\rho=x_1\rho x_2$.

(1) Multiplication. Two superoperators multiply as $XY\equiv x_1y_1\times y_2 x_2$.

(2) Transposition $(\T)$. The transposed operator $X^\T$ is related to $X$ by $X^\T\rho \equiv \rho X$, where
\begin{align}
    X^\T =x_2\times x_1\,,\qquad
    (cXY)^\T=cY^\T X^\T\,,\qquad
    (cX^\T)^\T=cX\,,
\end{align}
in which $c$ denotes a complex number. Symmetric superoperators satisfy $X^\T=X$.
	
(3) Adjunction $(\dg)$. The adjoint superoperator $X^\dg$ is related to $X$ by $\rho^\dg X^\dg\equiv (X\rho)^\dg$ where
\begin{align}
    X^\dg=x^\dg_1\times x^\dg_2\,,\qquad
    (cXY)^\dg=c^*Y^\dg X^\dg\,,\qquad
    (cX^\dg)^\dg=c^*X\,,
\end{align}
and $*$ denotes complex conjugate. Hermitian superoperators satisfy $X^\dg=X$.

(4) Association $(\sim)$. The associated operator $\widetilde{X}$ is defined through $\widetilde{X}\rho^\dg\equiv (X\rho)^\dg$. Association is the combined operations of transposition and adjunction,
\begin{align} \label{tilde}
    \widetilde{X}=(X^\dg)^\T=(X^\T)^\dg=x^\dg_2\times x^\dg_1\,,\qquad
    (cXY)\,\widetilde{\,}=c^*\widetilde{X}\widetilde{Y}\,,
    \qquad
    (c\widetilde{X})\,\widetilde{\,}=c^* X\,.
\end{align}
Following the definition of Ref.~\cite{Prigogine73}, superoperators are \textit{adjoint-symmetric} if $\widetilde{X}=X$.

\section{Time evolution operator}
\label{AppTimeEvo}

The solution to the equation of motion $\d\rho/\d t=-K(t)\rho$ is $\rho(t)=V(t)\rho(0)$, where the time evolution operator can be written in terms of the time-ordered products denoted by $\{\cdot\}_+$ \cite{Peskin95},
\begin{align} \label{rhot}
   V(t)&\equiv \left\{ \exp\left(-\int_{t_0}^t dt' K(t')\right)\right\}_+=1-\int_{t_0}^t dt_1 K(t_1) +\frac{(-1)^2}{2!}\int_{t_0}^t dt_1 \int_{t_0}^t dt_2 \{K(t_1) K(t_2)\}_++\cdots\,.
\end{align}
The evolution operators can also be presented as dynamical matrices \cite{Sudarshan61,Kraus83}.

Since the time evolution operators are transformation operators on the $\rho$, they must be adjoint-symmetric
\begin{align}
\widetilde{V}(t)=V(t)\,.
\end{align}
Using the second property association operation on the time-ordered products in \Eq{rhot}, we obtain
\begin{align}   \label{tmordtilde}
     \widetilde{V}(t)= \left\{ \exp\left(-\int_{t_0}^t dt' \Kt(t')\right)\right\}_+\,.
\end{align}
We conclude that the generator is also adjoint-symmetric \Eq{KtildeK}.

\section{Four dimensional matrix representation of the generators of symplectic group in four dimensions}
\label{AppMrepres}

The matrix representation of the generators can be obtained as follows. We introduce the column vector $\u{X}$ with four elements, $\u{X}_i, i=1, 2, 3, 4$, defined by
\begin{align} \label{Qv}
    \u{X} &\equiv
        \left(
          \begin{array}{c}
          A \\
          \At \\
          \Adg  \\
          \Atdg
          \end{array}
        \right)  \,.
\end{align}
We find that $[\u{X}_i,\u{X}_j]=\bbt_{ij}$, where $\bbt_{ij}$ are the components of the skew-symmetric matrix
\begin{align}   \label{beta}
    \bbt\equiv \left(\begin{array}{cc}0&\u{I}\\-\u{I}&0\end{array}\right),\qquad
        \u{I}=\left(\begin{array}{cc} 1&0\\0&1 \end{array} \right),
\end{align}

With a similarity transformation under $S(J)=\exp(\th J)$, where $J\in\mathbb{J}_0\cap\mathbb{J}_+\cap\mathbb{J}_-$, $X_i$ transforms as \cite{Simon88}
\begin{align}   \label{XS}
    \u{X}'_i&=S(J)\u{X}_iS^{-1}(J)=\sum_j \u{S}^{-1}_{ij}(J)\cdot\u{X}_j\,.
\end{align}	
In \Eq{XS},
\begin{align}
    \u{S}(J)\equiv\exp[\th \u{J}(J)]\,,
\end{align}	
where $\u{J}(J)$ is the four-dimensional (4D) matrix representation of $J$. For infinitesimal $\th$, we expand the exponential operators $S(J)$ and $\u{S}(J)$ on both sides of \Eq{XS} to first order in $\th$ to obtain
\begin{align}   \label{SJ}
    [J,\u{X}_i]=-\sum_j \u{J}_{ij}(J)\cdot\u{X}_j\,.
\end{align}
The matrix representation $\u{J}(J)$ of the generator $J$ can then be extracted from \Eq{SJ}. The results are
\begin{subequations}
\begin{align}   \label{L04d}
    \u{J}(iL_0) =\frac{i}{2} \left(\begin{array}{cc}
                        \s_3&0\\
                            0&-\s_3
          \end{array}\right),
\quad
        \u{J}(iM_1) = \frac{i}{2}\left(\begin{array}{cc}
            0&\s_3\\
            -\s_3&0
          \end{array}\right),
\quad
    \u{J}(iM_2) = \frac{1}{2}\left(\begin{array}{cc}
            0&\u{I}\\
            \u{I}&0
          \end{array}\right),
\quad
    \u{J}(O_0) = \frac{1}{2}\left(\begin{array}{cc}
            0&\s_1\\
            \s_1&0
          \end{array}\right),
\end{align}
\begin{align}
    \u{J}(O_+) = \frac{1}{2} \left(\begin{array}{cc}
            -\u{I}&\s_1\\
            -\s_1&\u{I}
          \end{array}\right),
\quad
    \u{J}(L_{1+}) = \frac{1}{2} \left(\begin{array}{cc}
            \s_1&-\u{I}\\
            \u{I}&-\s_1
          \end{array}\right),
\quad
    \u{J}(L_{2+}) = \frac{1}{2} \left(\begin{array}{cc}
            \s_2&i\s_3\\
            i\s_3&\s_2
          \end{array}\right),
\end{align}
\begin{align}   \label{L2-4d}
    \u{J}(O_-) = \frac{1}{2} \left(\begin{array}{cc}
            \u{I}&\s_1\\
            -\s_1&-\u{I}
          \end{array}\right),
\quad
     \u{J}(L_{1-}) = \frac{1}{2} \left(\begin{array}{cc}
            \s_1&\u{I}\\
            -\u{I}&-\s_1
          \end{array}\right),
\quad
    \u{J}(L_{2-}) = \frac{1}{2} \left(\begin{array}{cc}
                \s_2&-i\s_3\\
                -i\s_3&\s_2
          \end{array}\right),
\end{align}
\end{subequations}
where
\begin{align}   \label{sig}
    \s_1&=\left(\begin{array}{cc} 0&1\\1&0 \end{array} \right),&
    \s_2&=\left(\begin{array}{cc} 0&-i\\i&0 \end{array} \right),&
    \s_3&=\left(\begin{array}{cc} 1&0\\0&-1 \end{array} \right),
\end{align}
The generators satisfy
\begin{align}   \label{bJT}
   \bbt\cdot\u{J}(J)=\big[\bbt\cdot\u{J}(J)\big]^\T\,,
\end{align}
so that the commutation relation $[X'_i,X'_j]=\bt_{ij}$ is preserved. Moreover,
\begin{align}   \label{SSTbtSS}
    \u{S}^\T(J)\cdot\bbt\cdot\u{S}(J)=\bbt
\end{align}
is satisfied. Consequently, $S$ are elements of the symplectic group in four dimensions. \Eqs{L04d}-\eqref{L2-4d} obey the commutation relations in Table~\ref{commV}.
We can verify that they are complete,
\begin{align}   \label{Jcomplete}
    \sum_J \u{J}(J)\cdot\u{J}^\dg(J) =4 \u{I}\,,
\end{align}
where $\u{I}$ is the $4\times 4$ identity matrix, and orthogonal under the Hilbert-Schmidt norm,
\begin{align}   \label{HSnorm}
    \tr \big[\u{J}^\dg(J)\cdot\u{J}(J')\big]=c_J \del_{J,J'}\,,
\end{align}
where $c_J=1$ for $J\in\J_0$, and $c_J=2$ for $J\in \J_\pm$.

\section{Similarity transformation}
\label{AppTable}

The similarity transformation of an operator $O$ by $\exp(\al S)$ can be worked out by using the commutator bracket $[S,\cdot]$ repeatedly,
\begin{align}
    e^{\al S} Oe^{-\al S} &= O +\al [S,O] +\frac{\al^2}{2!}[S,[S, O]] +\cdots\,.
\end{align}

As a result, the similarity transformations of $K$, cf. \Eqs{Kprime} and \eqref{K}, under the various $S$ \eqref{UV} can be decomposed into the generators from the $\J_0$ and $\J_+$ spaces as follows.
\begin{align}   \label{URK}
    e^{\th iL_0} K e^{-\th iL_0}&=\h_0iL_0+ (\h_1 \cos\th + \h_2 \sin\th)iM_1+ (\h_2\cos\th -\h_1 \sin\th)iM_2+ g_0(O_0-I/2)\no
    &\quad+ g_+O_++ (g_1 \cos\th + g_2 \sin\th)L_{1+}+( g_2 \cos\th - g_1 \sin\th)L_{2+}\,,
\\  \label{U1K}
    e^{\phi iM_1} K  e^{-\phi iM_1}
    &=(\h_0\cosh\phi+\h_2\sinh\phi)iL_0+ \h_1iM_1+ (\h_2\cosh\phi +\h_0 \sinh\phi)iM_2 +g_0(O_0-I/2)\no
    &\quad+( g_+\cosh\phi+g_2\sinh\phi)O_++ g_1L_{1+}+( g_2 \cosh\phi + g_+ \sinh\phi)L_{2+}\,,
\\   \label{U2K}
     e^{\phi iM_2} K  e^{-\phi iM_2}
    &=(\h_0\cosh\phi-\h_1\sinh\phi)iL_0+(\h_1\cosh\phi -\h_0 \sinh\phi)iM_1+ \h_2iM_2  +g_0(O_0-I/2)\no
    &\quad+ (g_+\cosh\phi-g_1\sinh\phi)O_+ +(g_1\cosh\phi - g_+ \sinh\phi)L_{1+}+ g_2L_{2+}\,,
\\  \label{U0K}
    e^{\al O_0} K e^{-\al O_0}&=\h_0iL_0+ \h_1iM_1+ \h_2 iM_2  +g_0(O_0-I/2)+ e^\al g_+O_+  + e^\al g_1L_{1+}+ e^\al g_2 L_{2+}\,,
\\  \label{U+K}
    e^{\al O_+} K e^{-\al O_+}&=\h_0iL_0+ \h_1iM_1+ \h_2iM_2+g_0(O_0-I/2)+ (g_+ -\al g_0)O_+ +(g_1+\al\h_2)L_{1+}+( g_2-\al\h_1)L_{2+}\,,
\\  \label{U1+K}
   e^{\al L_{1+}} K e^{-\al L_{1+}}&=\h_0iL_0+ \h_1iM_1+ \h_2iM_2 +g_0 (O_0-I/2)+ (g_+ +\al\h_2)O_+  +(g_1- \al g_0)L_{1+}+( g_2+\al \h_0)L_{2+}\,,
\\  \label{U2+K}
    e^{\al L_{2+}} K e^{-\al L_{2+}}&=\h_0iL_0+ \h_1iM_1+ \h_2iM_2 +g_0(O_0-I/2)+ (g_+ -\al \h_1)O_+ +(g_1- \al \h_0)L_{1+}+( g_2-\al g_0)L_{2+}\,.
\end{align}

From the list, we learned that $\exp(\th iL_0), \exp(\th iM_1), \exp(\th iM_2)$ preserve the length $-h_0^2+h_1^2+h_2^2$ and $-g_+^2+g_1^2+g_2^2$ separately. $\exp(\al O_0)$ introduces a dilation factor to the coefficients of $O_+, L_{1+}, L_{2+}$, whereas $\exp(\al O_0), \exp(\al L_{1+}), \exp(\al L_{2+})$ cause a translation in the coefficients of $O_+, L_{1+}, L_{2+}$.

\section{Factorizable unitary transformation}
\label{AppRelU}

If a unitary superoperator $U$ is simultaneously adjoint-symmetric, using the fact that association consists of the combined operations of transposition and adjunction, cf.~\Eq{tilde}, we find that it also satisfies
\begin{align} \label{UT=U-1}
 U^\T=U^{-1}\,,
\end{align}
i.e., its generators is anti-symmetric. Factorizable unitary superoperators $u\times u^\dg$, where $u^\dg=u^{-1}$, for example, $u_\text{tot}$ that initiates the unitary transformation $u_\text{tot}\rho_\text{tot} u^\dg_\text{tot}$ in isolated quantum systems are naturally adjoint-symmetric. However, there exist non-factorizable unitary superoperators that do not satisfy \Eq{UT=U-1}, and hence are not adjoint-symmetric. This fact rules out some of the unitary superoperators that are not factorizable as valid transformation operators in the reduced space, such as $\exp(i\al O_+), \exp(i\al L_{1+})$ and $\exp(i\al L_{2+})$.

\section{Displacement operators}
\label{AppDop}

The displacement operators that generate the coherent states \cite{Klauder68,Gardiner} and the coherent thermofields \cite{Barnett85} are constructed by exponentiating operators linear in the four basic superoperators \eqref{AA}. There are two linear combinations of the superoperators that are adjoint-symmetric, namely, $z\Adg+z^*\Atdg$ and $yA+y^*\At$, where $y$ and $z$ are complex parameters. They produce factorizable superoperators upon exponentiations,
\begin{align}   \label{D1}
    D_1(z)\equiv \exp(z\Adg+z^*\Atdg)=e^{z\adg} \times (e^{z\adg})^\dg\,,\qquad
    D_2(y)\equiv \exp(yA+y^*\At)=e^{ya} \times (e^{ya})^\dg\,.
\end{align}
Under the similarity transformation, cf.~\Eq{Kprime}, we find that $D_1(z)$ shifts $A, \At$ into $A-z, \At-z^*$, respectively, while $\Adg, \Atdg$ remain invariant, whereas $D_2(y)$ shifts $\Adg, \Atdg$ into $\Adg+y, \Atdg+y^*$, respectively, while $A, \At$ remain invariant. Consequently, both the $D_1$ and $D_2$ preserve the commutation relations \eqref{commAA}. However, note that $D_1$ and $D_2$ are not unitary.

The displacement operators $D$ that generate the coherent states \cite{Klauder68,Gardiner} is obtained (up to a normalization factor) by taking the product
\begin{align}
    D(z)\equiv D_1(z)D_2(-z^*)&=e^{z\adg}e^{-z^*a}\times (e^{z\adg}e^{-z^*a})^\dg=e^{|z|^2} e^{z\adg-z^*a}\times (e^{z\adg-z^*a})^\dg\,,
\end{align}
which is now adjoint-symmetric, unitary and factorizable.

A generic generator of the reduced dynamics that annihilates the vacuum state is given by \Eq{KV}. This implies that the coherent states $D(z)|0;0\r$, are the stationary states of the reduced dynamics with the generic generator $D(z)K_VD^{-1}(z)$, whose explicit form is
\begin{align}
    D(z)K_VD^{-1}(z)=K_V+X+\widetilde{X}\,,\qquad
    X\equiv \frac{1}{2}\left[z(g_0+ih_0)+z^*(h_2+ih_1)\right](\At-\Adg)\,.
\end{align}

\section{The position representation}
\label{AppQr}

Using the definition $a\equiv (\xh + i \ph)/\sqrt{2}$, $\adg\equiv (\xh-i\ph)/\sqrt{2}$, where $\xh$ and $\ph=-i\d/\d x$ are dimensionless position and momentum operators, we find that
\begin{align} \label{AQr}
    \<x|A\rho|\xt\>=\frac{1}{\sqrt{2}}\left(x+\frac{\d}{\d x}\right)\<x|\rho|\xt\>\,,\qquad
    \<x|\Adg\rho|\xt\>=\frac{1}{\sqrt{2}}\left(x-\frac{\d}{\d x}\right)\<x|\rho|\xt\>\,,
\end{align}
where $x\equiv \sqrt{m\w_0}\,q$, $ \xt\equiv \sqrt{m\w_0}\,\widetilde{q}$, are dimensionless position coordinates of the oscillator in the bra- and ket-space, respectively, with units $\hbar=1$. $m$ and $\w_0$ are the mass and the natural frequency of the oscillator, respectively, whereas $q, \widetilde{q}$ are the position coordinates with the dimensions of length. The corresponding expressions of $\At,\Atdg$ operators can be obtained by replacing $x$ by $\xt$ in the round brackets of \Eq{AQr}, respectively. The center and relative coordinates are defined as
\begin{align}   \label{Qr}
    Q\equiv(x+\xt)/2\,, \qquad r\equiv x-\xt\,,
\end{align}
respectively.

In the position coordinates, the hermiticity condition $\rho^\dg=\rho$ takes the form $\<x|\rho^\dg|\xt\>=\<\xt|\rho|x\>^*=\<x|\rho|\xt\>$, or $\rho^*(Q,-r)=\rho(Q,r)$ \cite{Simon88}, where $*$ denotes complex conjugate. The association operation on operators $\widetilde{S}\rho^\dg=(S\rho)^\dg$ implies $\<x|\widetilde{S}\rho^\dg|\xt\>=\<\xt|S\rho|x\>^*$, or $\<x|\widetilde{S}|\xt\>=\<\xt|S|x\>^*$. Equivalently, we have $\widetilde{S}(Q,r)=S^*(Q,-r)$. Therefore, adjoint-symmetric operators satisfy $\widetilde{S}(Q,r)=S^*(Q,-r)=S(Q,r)$.

The generators of the reduced dynamics discussed in Section \ref{SecSymRD} take the following forms in the position coordinates,
\begin{align}   \label{KKLQr}
    &K_\KL(Q, r) = i\w_0\left(-\frac{\d^2}{\d Q\d r}+Qr\right)-\frac{\gam}{2}\left(Q\frac{\d}{\d Q}-r\frac{\d}{\d r}+1\right) -b\frac{\gam}{2}\left(\frac{\d^2}{\d Q^2}-r^2\right),  \\
    &K_\CL(Q, r) = i\w_0\left(-\frac{\d^2}{\d Q\d r}+Qr\right) +\gam r\frac{\d}{\d r} + b \gam r^2\,,\label{KCLQr}\\
    &K_\HPZ(Q, r) = i\w_0\left(-\frac{\d^2}{\d Q\d r}+Qr\right)+\gam r\frac{\d}{\d r} +b \gam  r^2 + i dr\frac{\d}{\d Q}\,. \label{KHPZQr}
\end{align}
We can now verify that \Eq{rhoequi} is the stationary state of the reduced dynamics.

\section{Expectation values}
\label{AppExpValue}

The expectation value of an observable $o$ is defined as \cite{Prigogine73}
\begin{align}   \label{expobserv}
    \<o\>_\rho=\tr (o^\dg\rho)\,.
\end{align}
Under the unitary transformation $o'\equiv Uo$ and $\rho'\equiv U\rho$, the expectation value is preserved,
\begin{align}
\<o'\>_{\rho'}=\tr (o'^\dg\rho')=\tr(o^\dg U^\dg U\rho)= \<o\>_\rho \,.
\end{align}
Note that this property is independent of the factorizability of $U$.

For transformation operators that are not unitary, the expectation values of the observables will not remain invariant. Since the transformation operators are time-independent, once we have normalized the transformed density operator, $\d\tr\rho'/\d t=0$ will be satisfied for all time. This is acceptable if the operators map the original dynamics into a physically inequivalent one. For example, in Section \ref{SecFormInv}, the transformed systems have different temperatures, whereas in Section \ref{SecBetwRedDyn}, the transformed systems have different dynamics from the original system.

In contrast, the star-unitary transformation introduced in Ref.~\cite{Prigogine73} is a transformation on the total system. The star-unitary transformation is not unitary and results in non-factorizable transformation on the subsystem \cite{Sungyun03}. It involves a combined operation of adjunction on operators (see \ref{AppOpSop}) and an inversion of the total Liouville operator. As a result, it leaves the expectation values of the operators invariant.


\providecommand{\noopsort}[1]{}\providecommand{\singleletter}[1]{#1}%

\end{document}